\documentclass[12pt]{article}
\usepackage{amssymb,amsmath,epsfig}
\begin{document}

\title{\bf Energy Contents of Some Well-Known Solutions in Teleparallel
Gravity}
\author{M. Sharif\thanks {msharif@math.pu.edu.pk} and Abdul
Jawad\thanks{jawadab181@yahoo.com}\\
Department of Mathematics, University of the Punjab,\\
Quaid-e-Azam Campus, Lahore-54590, Pakistan.}

\date{}

\maketitle
\begin{abstract}
In the context of teleparallel equivalent to General Relativity, we
study energy and its relevant quantities for some well-known black
hole solutions. For this purpose, we use the Hamiltonian approach
which gives reasonable and interesting results. We find that our
results of energy exactly coincide with several prescriptions in
General Relativity. This supports the claim that different
energy-momentum prescriptions can give identical results for a given
spacetime. We also evaluate energy-momentum flux of these solutions.
\end{abstract}


\section{Introduction}

One of the most interesting but challenging problems in Einstein
theory of General Relativity (GR) is the localization of energy.
This problem still needs a definite answer due to its
unusual nature and various viewpoints. Many renowned researchers
have devoted much attention to this intricate issue. The foremost
energy-momentum prescription was proposed by Einstein \cite{1}
himself. Following this, a large number of different energy-momentum
prescriptions \cite{2}-\cite{6} have been derived. However, many of these
are coordinate dependent, i.e., results will be meaningful only
if the calculations are performed in quasi-Cartesian coordinates.
M$\o$ller \cite{7} and Komar \cite{8}
tried to overcome this weakness and proposed a coordinate
independent energy-momentum prescription.

Penrose \cite{13} introduced coordinate independent prescription of
quasi-local mass. However, Bergqvist \cite{14} showed that no two
quasi-local mass definitions agreed for the
Reissner-Nordstr$\ddot{o}$m and Kerr spacetimes. Bernstein and Tod
\cite{15} explored the shortcomings of Penrose quasi-local mass
definition in handling the Kerr metric. Afterwards, Virbhadra
\cite{16,17} introduced the coincidence concept of different
energy-momentum prescriptions and gave a new spirit to this problem.
It has been investigated by many people \cite{18} that different energy-momentum
complexes demonstrate a high degree of consistency by yielding the
same and reasonable result for a given spacetime. Virbhadra and his
colleagues \cite{19,20} explored that Einstein, Landau-Lifshitz,
Papapetrou and Weinberg (ELLPW) prescriptions provide the same
results for any spacetime of Kerr-Schild class and more general
spacetimes if calculations are performed in Kerr-Schild Cartesian
coordinates. However, Sharif and Fatima \cite{21} found results
which did not support this viewpoint.

It has been found \cite{24,25} that telleparallel equivalent
to General Relativity (TEGR) provides more satisfactory solution of
the energy-momentum problem than does GR. This theory is based on
tetrad field initiated by M{\o}ller \cite{26}. Mikhail et al.
\cite{24} re-formulated M{\o}ller energy-momentum complex in this
alternative theory. Sharif and Amir \cite{30} found that energy for
the closed Friedmann model is consistent with GR. However, it was
concluded that energy-momentum prescriptions did not necessarily
provide consistent results in two theories, i.e., TEGR and GR.

Recent literature \cite{33}-\cite{36} shows keen interest in the
evaluation of energy-momentum by Lagrangian framework in TEGR. Maluf
et al. \cite{37} defined gravitational energy, momentum and angular
momentum by using this formulation \cite{38}. After that, many
people \cite{39} have used this procedure to evaluate energy and its
contents for different solutions. In a recent paper \cite{40}, we
have discussed energy and its related quantities for a class of
regular black holes coupled with non-linear electrodynamics source.

In this paper, this study is extended to evaluate energy and its
contents for some well-known black hole solutions by using the
Hamiltonian approach. The results will be compared to those found by
using different prescriptions in GR \cite{41}-\cite{44}. The paper
is organized as follows: In section \textbf{2}, we present the
formulation to evaluate gravitational energy, momentum, angular
momentum, gravitational and matter energy-momentum fluxes. Section
\textbf{3} contains brief discussion of some black hole solutions.
In section \textbf{4}, we evaluate energy, momentum and angular
momentum for these solutions. Section \textbf{5} is devoted to study
gravitational and matter energy-momentum fluxes. In the last
section, we present summary and discussion of the results obtained.

Following conventions are considered throughout the paper: Spacetime
indices $(\mu,\nu,\rho,...)$ and tangent space indices $(a,b,c,...)$
run from 0 to 3. Here $\mu=0,i$ and $a=(0),(i)$ denote time and
space indices respectively.

\section{Energy-Momentum and Hamiltonian Approach in Teleparallel Theory}

The Weitzenb\"{o}ck connection \cite{45} is defined in terms of
tetrad field ${e^a}_\mu$ as
\begin{equation}\label{1}
{\Gamma^\lambda}_{\mu\nu}={e_a}^\lambda\partial_\nu{e^a}_\mu
\end{equation}
and the torsion tensor
\begin{equation}\label{2}
{T^a}_{\mu\nu}=\partial_\mu{e^a}_\nu-\partial_\nu{e^a}_\mu.
\end{equation}
The Lagrangian density of the gravitational field in the presence of
matter \cite{38} is
\begin{equation}\label{3}
L\equiv-\kappa e\Sigma^{abc}T_{abc}-L_M,
\end{equation}
where $\kappa=1/16\pi,~e=det({e^a}_\mu)$ and the tensor
$\Sigma^{abc}$ is given by
\begin{equation}\label{4}
\Sigma^{abc}=\frac{1}{4}(T^{abc}+T^{bac}-T^{cab})
+\frac{1}{2}(\eta^{ac}T^b-\eta^{ab}T^c).
\end{equation}
Consequently, the field equations are
\begin{equation}\label{5}
e_{a\lambda}e_{b\mu}\partial_\nu(e\Sigma^{b\lambda\nu})
-e({\Sigma^{b\nu}}_a T_{b\nu\mu}-\frac{1}{4}e_{a\mu}
T_{bcd}\Sigma^{bcd})=\frac{1}{4\kappa}e T_{a\mu},\quad
\frac{\delta L_M}{\delta e^{a\mu}}=e T_{a\mu}.
\end{equation}
The total Hamiltonian density is defined as \cite{46}
\begin{equation}\label{6}
H(e_{ai},\Pi_{ai})=e_{a0}C^a+\alpha_{ik}\Gamma^{ik}+\beta_k\Gamma^k
+\partial_k(e_{a0}\Pi^{ak}),
\end{equation}
where $C^a,~\Gamma^{ik},~\Gamma^k$ and $\alpha_{ik},~\beta_k$
express primary constraints and Lagrangian multipliers
respectively.

The gravitational energy-momentum over an arbitrary volume $V$ is
\begin{equation}\label{7}
P^a=-\int_V d^3 x\partial_i\Pi^{ai},
\end{equation}
where $-\partial_i\Pi^{ai}=\partial_i(4\kappa e\Sigma^{a0i})$
represents the \textit{energy-momentum density} \cite{37}. The
above expression can be written in terms of fluxes as
\begin{equation}\label{8}
\frac{dP^a}{dt}=-\Phi^a_g-\Phi^a_m,
\end{equation}
where
\begin{equation}\label{9}
\Phi^a_g=\int_S dS_j\phi^{aj},\quad \Phi^a_m=\int_S dS_j(e{e^a}_\mu
T^{j\mu})
\end{equation}
represent the $a$ component of the \textit{gravitational
energy-momentum flux} and \textit{matter energy-momentum flux}
\cite{47} while $S$ is the spatial boundary of the volume $V$. The
quantity $\phi^{aj}$ describe the $a$ component of the gravitational
energy-momentum flux density in $j$ direction and its expression is
\begin{equation}\label{10}
\phi^{aj}=\kappa e e^{a\mu}(4\Sigma^{bcj}T_{bc\mu}
-\delta^j_\mu\Sigma^{bcd}T_{bcd}).
\end{equation}
The total angular momentum \cite{48} is
\begin{equation}\label{11}
M^{ik}=2\kappa\int_V d^3x
e[-g^{im}g^{kj}{T^0}_{mj}+(g^{im}g^{0k}-g^{km}g^{0i}){T^j}_{mj}].
\end{equation}

\section{Black Hole Solutions}

The generalized form of black hole solutions is given by
\begin{equation}\label{12}
ds^2=-Fdt^2+F^{-1}dr^2+r^2d\theta^2+r^2{\sin^2\theta}d\phi^2,
\end{equation}
where $F= (1-2\frac{M(r)}{r})$. This metric can be reduced to some
well-known black holes under special choices of $M(r)$. Some of them
are given below.

\subsection{de Sitter-Schwarzschild Black Hole}

Dymnikova \cite{49} found a regular black hole solution in de Sitter
form which exhibits Schwarzschild like behavior by
replacing its singularity with de Sitter core. The line element is
found for
\begin{equation}\label{13}
M_1(r)=m(1-\exp(-\frac{r^3}{r^3_{*}})),
\end{equation}
where $r^3_{*}=2mr^2_0,~r^2_0=\frac{3}{\Lambda}$. It behaves
asymptotically Schwarzschild and de Sitter solutions when $r
\rightarrow \infty$ and $r \rightarrow 0$ respectively. Here
$\Xi_1=0$.

\subsection{Regular Black Hole with Cosmological Constant}

Mo Wen-Juan et al. \cite{50} introduced a class of regular black
hole solution with cosmological constant $\Lambda$ in non-linear
electrodynamics. The solution is given in the metric form with
\begin{equation}\label{14}
M_2(r)=(\frac{mr^{3}}{(r^2+q^2)^{3/2}}-
\frac{q^2r^{3}}{2(r^2+q^2)^{2}}+\frac{\Lambda r^3}{6}),
\end{equation}
where $m,~q$ and $\Lambda$ represent mass, electric charge and
cosmological constant respectively. For $\Lambda=0$,
the solution reduces to the case discussed by Ayon-Beato and Garcia
\cite{51} in which they have shown that the field strength and all
curvature invariants are regular everywhere. The appearance of
cosmological constant does not destroy the regularity of the
solution. The associated electric field strength $\Xi_2$ is given by
\begin{equation}\label{15}
\Xi_2=qr^4(\frac{r^2-5q^2}{(r^2+q^2)^4}+\frac{15m}{2(r^2+q^2)^{7/2}}).
\end{equation}
For $m=0=q$, the solution represents the dS ($\Lambda> 0$) or AdS
($\Lambda < 0$) space. The solution behaves asymptotically
Reissner-Nordstr$\ddot{o}$m black hole in dS/AdS space depending on
the sign of the cosmological constant $\Lambda$ as
\begin{equation}\label{16}
F=1-\frac{2m}{r}+\frac{q^2}{r^2}+O(\frac{1}{r^3})-\frac{\Lambda~r^2}{3}.
\end{equation}
This class reduces to the Schwarzschild solution for $q=0=\Lambda$.

\subsection{Bardeen Regular Black Hole}

Ayon-Beato and Garcia \cite{52} gave physical interpretation of
Bardeen model \cite{53} by showing that charge associated with it
acts as a magnetic monopole charge. This is described by the metric
with
\begin{equation}\label{17}
M_3(r)=\frac{mr^3}{(r^2+e^2)^{3/2}}
\end{equation}
and the associated magnetic field strength is given by
\begin{equation}\label{18}
\Xi_3=\frac{e^2}{2r^4}.
\end{equation}
Here $m$ and $e$ stand for mass and monopole charge of a
self-gravitating magnetic field of non-linear electrodynamics source
respectively. This solution exhibits black hole behavior for $e^2
\leq (16/27)m^2$. The curvature invariants corresponding to this
solution are regular everywhere. It behaves asymptotically as
\begin{equation}\label{19}
F=1-\frac{2m}{r}+\frac{3me^2}{r^3}+O(\frac{1}{r^5}).
\end{equation}
For $e=0$, the solution reduces to the Schwarzschild spacetime.

\subsection{Dyadosphere of a Reissner-Nordstr$\ddot{o}$m Black Hole}

According to Ruffini \cite{54}, the dyadosphere is defined as the
region outside the horizon of an electromagnetic black hole where
the electromagnetic field is stronger than the well-known
Heisenbeg-Euler critical value for electron-positron pair production
\begin{equation}\label{20}
\varepsilon_{cr}=\frac{m^2_{\tilde{e}}c^3}{\tilde{e}\hbar}.
\end{equation}
Here $m_{\tilde{e}}$ and $\tilde{e}$ play the role of mass and
charge of an electron. This concept was introduced to
explain gamma ray bursts. The dyadosphere region for
Reissner-Nordstr$\ddot{o}$n black hole is described by the radial
interval $r_+\leq r\leq r_{ds}$, where $r_+$ (called horizon of
black hole) and $r_{ds}$ represent the inner and outer radii of the
dyadosphere. The expressions for $r_+$ and $r_{ds}$ are given by
\begin{eqnarray}
\label{21}r_+ &=&\frac{Gm}{c^2}(1\pm\sqrt{1-\frac{q^2}{Gm^2}}),\\
\label{22}r_{ds}&=&\sqrt{(\frac{\hbar}{m_{\tilde{e}}c})(\frac{Gm}{c^2})
(\frac{m_{pl}}{m_{\tilde{e}}})(\frac{\tilde{e}}{q_{pl}})(\frac{q}
{\sqrt{G}m})}
\end{eqnarray}
respectively. Here $m,~q,~m_{pl}=\sqrt{\frac{\hbar c}{G}}$ and
$q_{pl}=\sqrt{\hbar c}$ denote mass, charge, Planck mass and Planck
charge respectively. The electron-positron pair production processes
occur over the whole region of  dyadosphere and hence the total
energy confined in the dyadosphere is given by \cite{55}
\begin{equation*}
E_{dya}=\frac{q^2}{2r_+}(1-\frac{r_+}{r_{ds}})(1-\frac{r^2_+}{r^2_{ds}}).
\end{equation*}
De Lorenci et al. \cite{56} obtained Reissner-Nordstr$\ddot{o}$m
black hole in the dyadosphere form with
\begin{equation}\label{23}
M_4(r)=m-\frac{q^2}{2r}+\frac{\sigma q^4}{10r^5}.
\end{equation}
The Reissner-Nordstr$\ddot{o}$n solution arises for the vanishing of
$\sigma$. The corresponding electric field source is
\begin{equation}\label{24}
\Xi_4=-\frac{q}{r^2}+4\frac{\sigma q^3}{r^6}.
\end{equation}

\section{Energy, Momentum and Angular Momentum}

The tetrad components associated to (\ref{12}) can be obtained by
using the procedure of \cite{31} as
\begin{equation}\label{25}
{e^a}_\mu(r,\theta,\phi) =\left(\begin{array}{cccc}
\sqrt{F} & 0 & 0 & 0 \\
0 & \frac{1}{\sqrt{F}}\cos\phi\sin\theta & r\cos\phi\cos\theta & -r\sin\phi\sin\theta \\
0 &  \frac{1}{\sqrt{F}}\sin\phi\sin\theta & r\sin\phi\cos\theta & r\cos\phi\sin\theta \\
0 & \frac{1}{\sqrt{F}}\cos\theta & -r\sin\theta & 0\\
\end{array}
\right)
\end{equation}
with $e=det({e^a}_\mu)=r^2\sin\theta$. The non-vanishing components
of torsion tensor are found by substituting the tetrad
components in Eq. (\ref{2}), i.e.
\begin{eqnarray}\label{26}
T_{(0)01}&=&\dot{\sqrt{F}},\quad
\quad T_{(1)12}=(1-\frac{1}{\sqrt{F}})\cos\theta\cos\phi,\nonumber\\
T_{(1)13}&=&-(1-\frac{1}{\sqrt{F}})\sin\theta\sin\phi,\quad
T_{(2)12}=(1-\frac{1}{\sqrt{F}})\cos\theta\sin\phi,\nonumber\\
T_{(2)13}&=&(1-\frac{1}{\sqrt{F}})\sin\theta\cos\phi,\quad
T_{(3)12}=-(1-\frac{1}{\sqrt{F}})\sin\theta.
\end{eqnarray}
The corresponding non-zero components of $T_{\lambda\mu\nu}=
{e^a}_\lambda T_{a\mu\nu}$ will become
\begin{eqnarray}\label{27}
T_{001}=\sqrt{F}\dot{\sqrt{F}},\quad
T_{212}=r(1-\frac{1}{\sqrt{F}}),\quad
T_{313}=r\sin^2\theta(1-\frac{1}{\sqrt{F}}),
\end{eqnarray}
where dot indicates derivative with respect to $r$. In view of
Eq.(\ref{4}), the energy density corresponding to (\ref{12}) is
\begin{equation}\label{28}
-\partial_i\Pi^{(0)i}=4\kappa r\partial_1(\sin\theta(1-\sqrt{F})).
\end{equation}
Consequently, the energy turns out to be
\begin{eqnarray}\label{29}
P^{(0)}=E&=&r[1-\sqrt{F}],\nonumber\\
E&=&r[1-\sqrt{1-\frac{2M(r)}{r}}].
\end{eqnarray}
Making use of the binomial expansion with $r\gg M(r)$, it takes the
form
\begin{equation}\label{30}
E{\approx}M(r).
\end{equation}
For different black hole solutions, we have the following form of energy:\\
\begin{itemize}
\item For \textbf{de Sitter-Schwarzschild black hole}, inserting Eq.(\ref{13})
in the above equation, it follows that
\begin{equation}\label{31}
E_1=m(1-\exp(-\frac{r^3}{r^2_{*}})).
\end{equation}
This energy distribution is the same as calculated by Dymnikova
\cite{49} using the standard formula for mass of the de
Sitter-Schwarzschild solution, i.e.
\begin{equation}\label{32}
m(r)=\int^r_0T^0_0d^3x = R_g(r)/ 2.
\end{equation}
This also coincides with the results of Yang and Radinschi \cite{41}
evaluated by using Einstein, Weinberg and Tolman prescriptions
respectively, i.e.
\begin{equation}\label{33}
E_1=E_E=E_W=E_T=m(r)=R_g(r)/ 2=m(1-\exp(-\frac{r^3}{r^2_{*}})).
\end{equation}
We see that the energy $E_1$ vanishes at $r=0$ while for $r
\rightarrow \infty$ it is $m$. Thus the total energy is given by the parameter
$m$ which is the same as the $\textbf{ADM}$ mass for this spacetime.
Also, $E(r)> 0$ when $0 \leq r < \infty$.\\
\item For the \textbf{Regular black hole with cosmological constant}, we
have
\begin{equation}\label{34}
E_2=\frac{mr^{3}}{(r^2+q^2)^{3/2}}-
\frac{q^2r^{3}}{2(r^2+q^2)^{2}}+\frac{\Lambda r^3}{6}.
\end{equation}
When there is no cosmological constant, it turns out to be
\begin{equation}\label{35}
E_2=\frac{m}{(1+\frac{q^2}{r^2})^{3/2}}-
\frac{q^2}{2r(1+\frac{q^2}{r^2})^2}.
\end{equation}
This expression is exactly the same as found by Yang et al.
\cite{42} in GR using Einstein, Weinberg energy-momentum
prescriptions but different from M$\o$ller complex. It is also
discussed by the same authors \cite{57} as a special case. After
applying power series expansion, the above expression will become
\begin{equation}\label{36}
E_2=m-\frac{q^2}{2r}-\frac{3mq^2}{2r^2}+\frac{q^4}{r^3}
+\frac{15mq^4}{8r^4}-\frac{3q^6}{2r^5}+O(\frac{1}{r^6}).
\end{equation}
It can also be written as
\begin{equation}\label{37}
E_2=E_{Tod}-\frac{3mq^2}{2r^2}+\frac{q^4}{r^3}
+\frac{15mq^4}{8r^4}-\frac{3q^6}{2r^5}+O(\frac{1}{r^6}),
\end{equation}
where $E_{Tod}$ represents the energy computed by Tod \cite{58} for
Reissner-Nordstr$\ddot{o}$m solution by using Penrose quasi-local
mass definition.\\
\item The energy of \textbf{Bardeen regular black hole} turn sout to be
\begin{equation}\label{38}
E_3=\frac{mr^3}{(r^2+e^2)^{3/2}}.
\end{equation}
This is the same as found by Sharif \cite{43}
\begin{equation}\label{39}
E_{ELLPW}=\frac{mr^3}{(r^2+e^2)^{3/2}}.
\end{equation}
He also employed the M$\o$ller energy-momentum prescription and
concluded that it also coincides with ELLPW prescriptions at large
distances. This reduces to the energy of Schwarzschild solution for
$e=0$.\\
\item \textbf{The dyadosphere of a Reissner-Nordstr$\ddot{o}$m} black hole
has the following energy distribution
\begin{equation}\label{40}
E_4=m-\frac{q^2}{2r}+\frac{\sigma q^4}{10r^5}.
\end{equation}
This result exactly agrees with ELLPW energy-momentum prescriptions
obtained by Xulu \cite{44} and slightly different from M$\o$ller
prescription \cite{59}. For $\sigma=0$, this corresponds to
Reissner-Nordstr$\ddot{o}$m black hole. The energy-momentum
prescriptions ELLPWM give consistent result for this special case
and reduces to the mass of Schwarzschild solution for $q=0$.
\end{itemize}

We would like to mention here that the momentum and angular momentum turn out to
be constant for all these solutions.

\section{Energy-Momentum Flux}

The gravitational energy flux becomes constant due to vanishing of all the components
of gravitational energy flux density $\phi^{(0)j}$, i.e. for $a=0$,
we have $\Phi^{(0)}_g=\textmd{constant}$. We carry out the
calculations for momentum flux density and obtain the following
components
\begin{eqnarray}\label{42}
\phi^{(1)1}&=&2\kappa\sin^2\theta\cos\phi(\sqrt{F}(\sqrt{F}-1)^2),\nonumber\\
\phi^{(1)2}&=&2\kappa\sin\theta\cos\theta\cos\phi(\sqrt{F}\dot{)}(\sqrt{F}-1),\nonumber\\
\phi^{(1)3}&=&-2\kappa\sin\phi(\sqrt{F}\dot{)}(\sqrt{F}-1),\nonumber\\
\phi^{(2)1}&=&2\kappa\sin^2\theta\sin\phi(\sqrt{F}(\sqrt{F}-1)^2),\nonumber\\
\phi^{(2)2}&=&2\kappa\sin\theta\cos\theta\sin\phi(\sqrt{F}\dot{)}(\sqrt{F}-1),\nonumber\\
\phi^{(2)3}&=&2\kappa\cos\phi(\sqrt{F}\dot{)}(\sqrt{F}-1),\nonumber\\
\phi^{(3)1}&=&2\kappa\sin\theta\cos\theta(\sqrt{F}(\sqrt{F}-1)^2),\nonumber\\
\phi^{(3)2}&=&-2\kappa\sin^2\theta(\sqrt{F}\dot{)}(\sqrt{F}-1),\nonumber\\
\phi^{(3)3}&=&0.
\end{eqnarray}
By replacing $a=i=1,2,3$ in Eq. (\ref{9}), we get the momentum
flux as follows
\begin{eqnarray}\label{43}
\Phi^{(1)}_g&=&-2\kappa\pi\sin\phi(\frac{F}{2}-\sqrt{F})+const,\nonumber\\
\Phi^{(2)}_g&=&2\kappa\pi\cos\phi(\frac{F}{2}-\sqrt{F})+const,\nonumber\\
\Phi^{(3)}_g&=&-4\kappa\pi\sin^2\theta(\frac{F}{2}-\sqrt{F})+const.
\end{eqnarray}
These turn out to be constant when we apply the condition $r\gg
M(r)$, i.e.
\begin{equation*}
\frac{F}{2}-\sqrt{F}=(1/2-M(r)/r)-\sqrt{(1-2M(r)/r)}\approx-\frac{1}{2}.
\end{equation*}
Consequently, the above expressions reduce to
\begin{eqnarray}\label{44}
\Phi^{(1)}_g&=&\kappa\pi\sin\phi+const,\nonumber\\
\Phi^{(2)}_g&=&-\kappa\pi\cos\phi+const,\nonumber\\
\Phi^{(3)}_g&=&2\kappa\pi\sin^2\theta+const.
\end{eqnarray}
Thus the components of gravitational momentum flux are free of
parameters like $m,~q$ and $\Lambda$. They only depend upon
spherical coordinates $\theta$ and $\phi$.

Now we calculate matter energy-momentum flux for de
Sitter-Schwarzschild black hole which requires the non-zero
components of the energy-momentum tensor
\begin{eqnarray}\label{45}
T^{00}&=&\frac{\Lambda e^\frac{-\Lambda
r^3}{6m}}{\kappa(1-\frac{2m}{r}+
\frac{2me^{\frac{-\Lambda r^3}{6m}}}{r})},\nonumber\\
T^{11}&=&{\frac{\Lambda}{\kappa}e^\frac{-\Lambda
r^3}{6m}(1-\frac{2m}{r}+
\frac{2me^\frac{-\Lambda r^3}{6m}}{r})},\nonumber\\
T^{22}&=&\frac{\Lambda}{2\kappa r^2}(2-\frac{r^3}{2m})e^{\frac{-\Lambda r^3}{6m}},\nonumber\\
T^{33}&=&\frac{\Lambda}{2\kappa
r^2\sin^2\theta}(2-\frac{r^3}{2m})e^{\frac{-\Lambda r^3}{6m}}.
\end{eqnarray}
The matter energy-momentum flux turns out to be
\begin{eqnarray}\label{46}
\Phi^{(0)}_m&=&const,\nonumber\\
\Phi^{(1)}_m&=&-\frac{\pi r^2\Lambda}{2\kappa}e^\frac{-\Lambda r^3}{6m}\cos\phi+const,\nonumber\\
\Phi^{(2)}_m&=&\frac{\pi r^2\Lambda}{2\kappa}e^\frac{-\Lambda r^3}{6m}\sin\phi+const,\nonumber\\
\Phi^{(3)}_m&=&-\frac{1}{\kappa}\pi r^2\Lambda\sin^2\theta
e^\frac{-\Lambda r^3}{6m}+const.
\end{eqnarray}
The matter energy-momentum flux corresponding to charged black hole
solutions can be evaluated by using the electromagnetic
energy-momentum tensor. Its non-zero components are
\begin{eqnarray}\label{47}
T^{00}&=&-\frac{\Xi^{2}}{8{\pi}F},\quad T^{11}=\frac{F\Xi^2}{8\pi},\nonumber\\
T^{22}&=&-\frac{\Xi^{2}}{8{\pi}r^2},\quad
T^{33}=-\frac{\Xi^{2}}{8{\pi}r^2\sin^2\theta}.
\end{eqnarray}
Here $\Xi$ is the electric field related to each charged black hole
solution. The energy flux of matter becomes constant and the
matter momentum flux for the charged black hole solutions are give as
follows:
\begin{itemize}
\item The matter flux for regular black hole solution with cosmological
constant is evaluated by substituting the electric field $\Xi_2$ in
Eq.(\ref{47}), i.e.
\begin{eqnarray}\label{48}
\Phi^{(1)}_m&=&\frac{1}{8}\sin\phi\int(r\Xi_2^2)dr+const,\nonumber\\
\Phi^{(2)}_m&=&-\frac{1}{8}\cos\phi\int(r\Xi_2^2)dr+const,\nonumber\\
\Phi^{(3)}_m&=&\frac{1}{4}\sin^{2}\theta\int(r\Xi_2^2)dr+const.
\end{eqnarray}
\item The matter flux for Bardeen regular black hole takes the following form
\begin{eqnarray}\label{49}
\Phi^{(1)}_m&=&\frac{-7e^4}{32r^8}\sin\phi+const,\nonumber\\
\Phi^{(2)}_m&=&\frac{7e^4}{32r^8}\cos\phi+const,\nonumber\\
\Phi^{(3)}_m&=&\frac{-7e^4}{16r^8}\sin^{2}\theta+const.
\end{eqnarray}
\item The dyadosphere of a Reissner-Nordstr$\ddot{o}$m black
hole has the matter flux
\begin{eqnarray}\label{50}
\Phi^{(1)}_m&=&\frac{-q^2}{8r^4}\sin\phi[3+\frac{176\sigma^2q^4}{r^8}
-\frac{56\sigma^2q^2}{r^4}]+const,\nonumber\\
\Phi^{(2)}_m&=&\frac{q^2}{8r^4}\cos\phi[3+\frac{176\sigma^2q^4}{r^8}
-\frac{56\sigma^2q^2}{r^4}]+const,\nonumber\\
\Phi^{(3)}_m&=&\frac{-q^2}{4r^4}\sin^{2}\theta[3+\frac{176\sigma^2q^4}{r^8}
-\frac{56\sigma^2q^2}{r^4}]+const.
\end{eqnarray}
\end{itemize}
The values of $\Phi^a_g$ and $\Phi^a_m$ represent the transfer of
gravitational and matter energy-momentum respectively.

\section{Summary and Discussion}

The debate of energy localization has generated a great deal of
interest for a number of scientists in GR and TEGR, but could not provide
a unique answer. In the current work, we
have computed the gravitational energy and its relevant quantities
like momentum, angular momentum, gravitational and matter
energy-momentum fluxes. We have investigated these quantities for
four well-known black hole solutions, i.e., de Sitter-Schwarzschild
black hole, regular black hole solution with cosmological constant, Bardeen
regular black hole and dyadosphere of a charged black hole. For this purpose, we have
used Hamiltonian approach in the realm of TEGR. It is worthwhile to
mention here that our results for energy distribution exactly
coincide with those evaluated by different authors using different
prescriptions in GR. These are given by Eqs.(\ref{31}), (\ref{34}),
(\ref{38}) and (\ref{40}). It is also interesting to note that these expressions
reduce to $\textbf{ADM}$ mass and this result also corresponds to
the Schwarzschild solution. Our results support the idea that the
energy-momentum complexes can give the same result for a given
spacetime. The momentum and angular momentum for these solutions
become constant.

The gravitational and matter energy-momentum flux have also been
evaluated for these black hole solutions. We find that the
gravitational and matter energy flux vanish while the components of
gravitational momentum flux become constant in the asymptotic
region. This indicates that the flow is uniform in that region and
occurs in the $\theta$ and $\phi$ directions. Moreover, the
components of matter flux associated to the de Sitter-schwarzschild
black hole depends on $\Lambda,~\theta$ and $\phi$. The matter flow
shows inward and outward falling for the variation of these
parameters. The components of matter flux corresponding to the
charged black holes are given in Eqs.(\ref{48})-(\ref{50}). The
matter flux of the de Sitter-Schwarzschild and charged black holes
vanish for $\Lambda=0$ and $q=0$ respectively.

\end{document}